\documentclass[a4paper,11pt]{article}
\usepackage{pos}

\renewcommand{\logo}{\relax}

\title{A novel phenomenological approach to total charm cross section measurements at the LHC}

\author*[a]{Yewon Yang}
\author[a]{Achim Geiser}

\affiliation[a]{Deutsches Elektronen-Synchrotron DESY, \\Notkestr. 85, 22607 Hamburg, Germany }

\emailAdd{yewon.yang@desy.de}
\emailAdd{achim.geiser@desy.de}

\abstract{
Measuring the total charm cross section is important for the comparison to theoretical predictions of the highest precision available for charm today, which are completely known up to NNLO QCD for the total inclusive cross sections. These are also independent of charm fragmentation, while practical measurements of charm hadrons in a fiducial phase space are not. Recently the LHC experiments have reported non-universality of charm fragmentation, which shows that e.g. charm baryon-to-meson ratios are not universal in different collision systems, and that the related production fractions also depend on transverse momentum. This breaks the charm fragmentation universality that was assumed until recently for the extrapolation of experimental measurements to the full total charm cross section phase space. A proposal is made how to address this non-universality in a data driven way without the need to implement any particular non-universal fragmentation model. As a practical example, this method is applied to the extrapolation of published LHC measurements of $D^0$ production at $\sqrt{s}=5$ TeV to the corresponding total charm cross section, which fully accounts for charm fragmentation non-universality for the first time. The result, $8.43 ^{+1.05}_{-1.16}(\text{total})$ mb, differs substantially from the one assuming charm fragmentation universality, but still compares well to theoretical QCD predictions up to NNLO.
}

\FullConference{The European Physical Society Conference on High Energy Physics (EPS-HEP2023)\\
 21-25 August 2023\\
Hamburg, Germany\\}

\begin{document}

\begin{flushright}
DESY-23-183\\
\end{flushright}

\maketitle

\section{Introduction}

The theory of Quantum-Chromo-Dynamics (QCD) is a well established part of the Standard Model, which describes many of the processes occurring in $pp$ collisions at LHC. Predictions for charm production are in particular challenging since, due to the closeness of the charm mass to $\Lambda_{\text{QCD}}$, the convergence of the perturbative series is slow resulting in large theoretical uncertainties. Charm measurements thus test QCD in the transition region of the perturbative and non-perturbative regimes. Measuring the total charm-quark pair cross section ($\sigma_{c\bar{c}}$) without any cuts on phase space is particularly important since for charm the corresponding theoretical predictions are the only ones available at next-to-next-to-leading order (NNLO), and do furthermore not depend on charm fragmentation. For such measurements, differential distributions measured in limited kinematic ranges and for a restricted set of final states need to be extrapolated to the total cross section, under certain theoretical assumptions. 

Using the QCD factorization theorem, open heavy-quark hadron production in $pp$ collisions is traditionally expressed as a convolution of the parton distribution functions (PDFs) $f_i$ and $f_j$ for the initial partons $i$ and $j$, the partonic cross section $\hat{\sigma}_{ij\rightarrow Q\bar{Q}}$ for the production of a heavy-quark pair $Q\bar{Q}$, and the non-perturbative fragmentation function $D_{Q\rightarrow H_Q}^{\text{NP}}$ for the (up to QCD evolution) universal fragmentation of one of the two heavy quarks into a particular open heavy-quark hadron $H_Q$:
\begin{equation} \label{eq:hqhXsec}
 d\sigma_{H_Q} \propto d\sigma_{pp \rightarrow Q\bar{Q}} \otimes D_{Q\rightarrow H_Q}^{\text{NP}}, \quad d\sigma_{pp \rightarrow Q\bar{Q}} = f_i f_j \otimes d\hat{\sigma}_{ij\rightarrow Q\bar{Q}},
\end{equation}
where the symbol $\otimes$ indicates the convolution and $d\sigma$ stands for a differential cross section distribution. The partonic cross section can be calculated as a truncated expansion of the QCD perturbative series in terms of powers of the strong coupling constant $\alpha_s$, optionally supplemented by the all-order resummation of logarithmic terms. The highest order currently available calculations of single inclusive transverse momentum ($p_T$) and (pseudo-)rapidity (($\eta$)$y$) distributions for charm are given by NLO+NLL calculations (FONLL\cite{fonll1, fonll2}). The non-perturbative PDFs and fragmentation functions, on the other hand, should be determined from experiments. Until recently, charm fragmentation was assumed to be universal, such that the fragmentation input was extracted mostly based on $e^+e^-$/$ep$ data, and applied also to $pp$ collisions. Especially, charm fragmentation fractions, which represent the integrated probability of charm to fragment into a particular hadron state, have been measured precisely from $e^+e^-$ and $ep$ collisions, assuming fragmentation universality, and no significant difference has been reported between the two (e.g., \cite{fragfrac_ep}). 

\begin{figure}
 \begin{center}
  \includegraphics[width=0.3\textwidth]{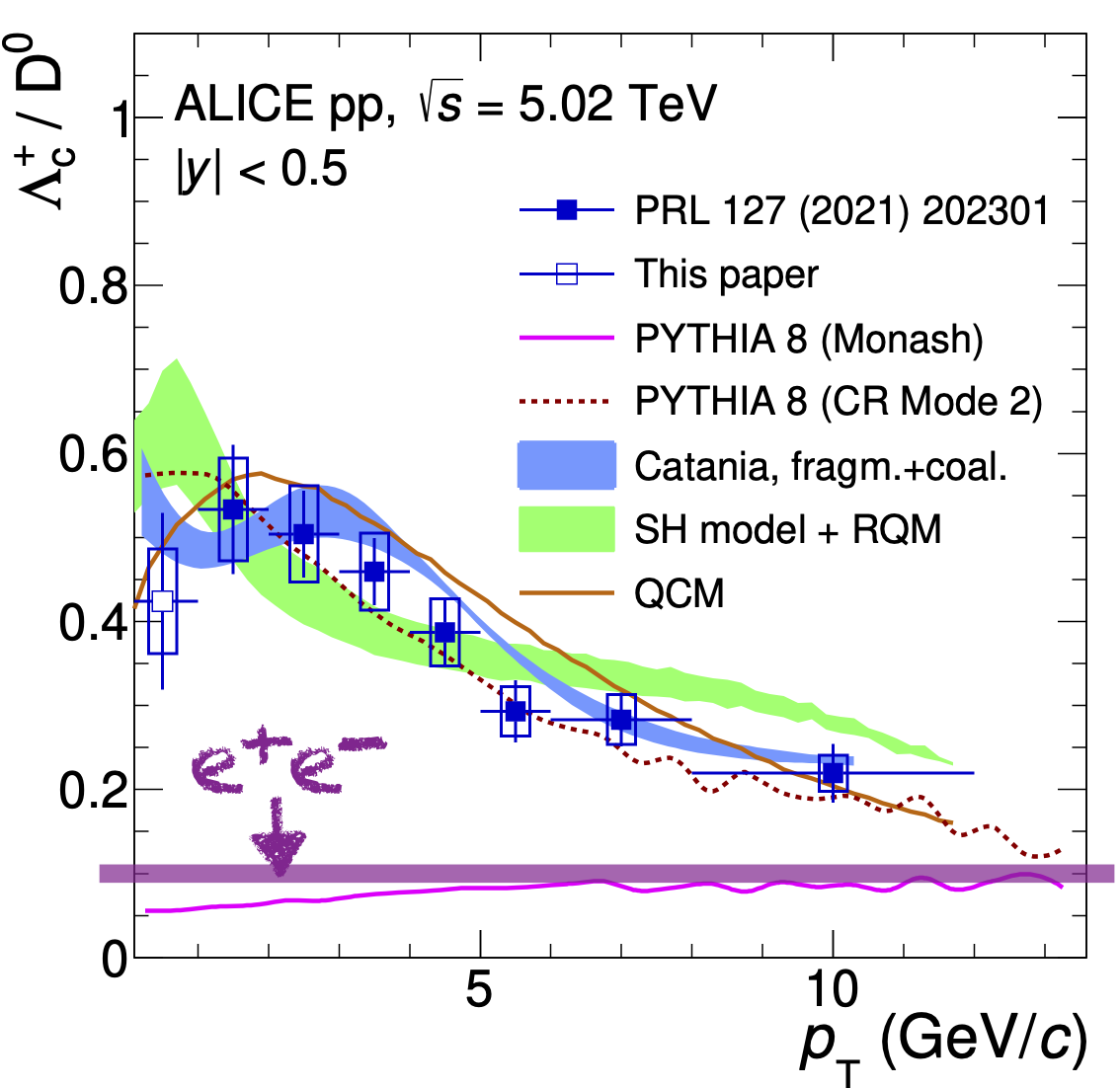}
  \includegraphics[width=0.3\textwidth]{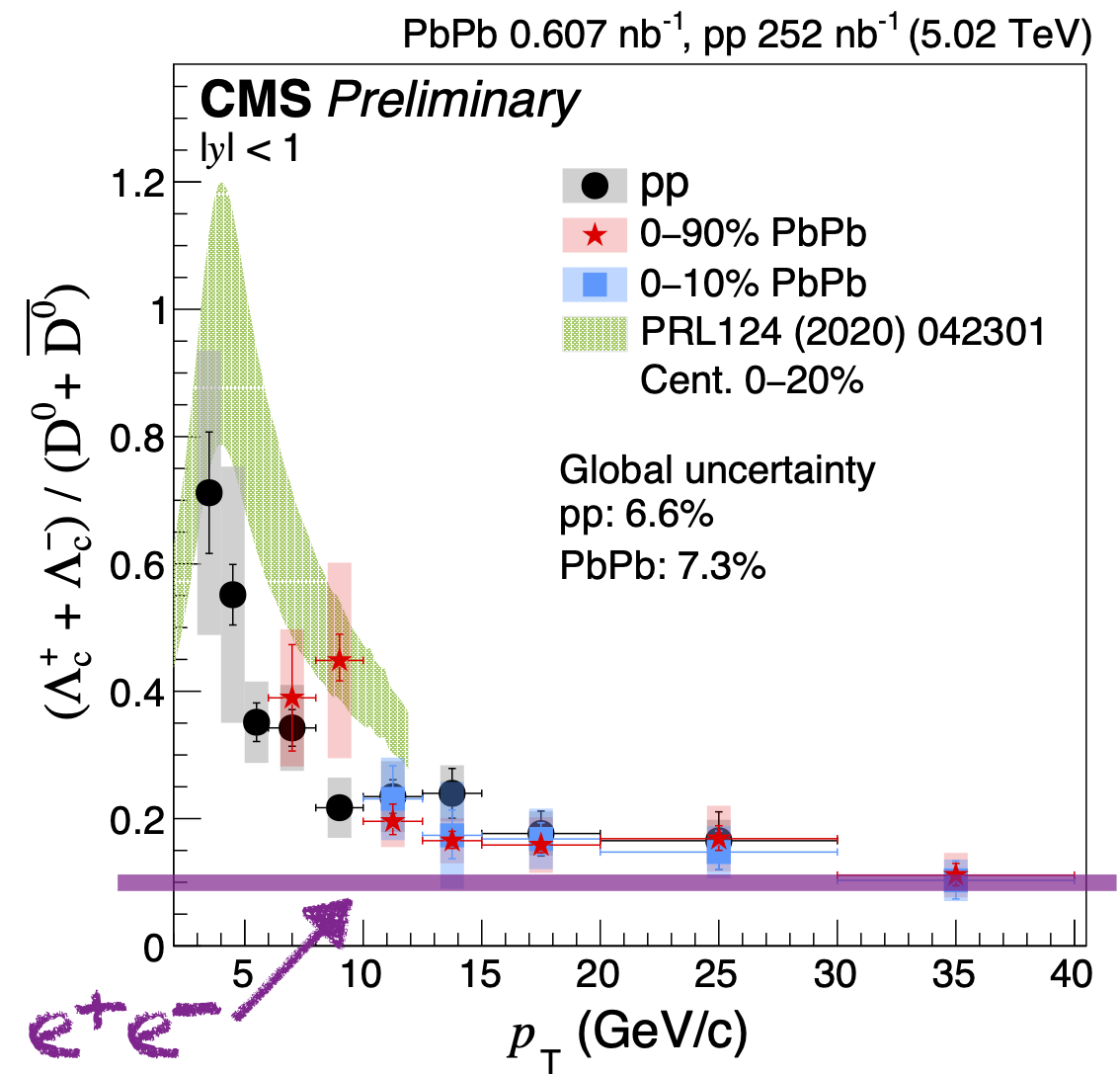}
  \includegraphics[width=0.35\textwidth]{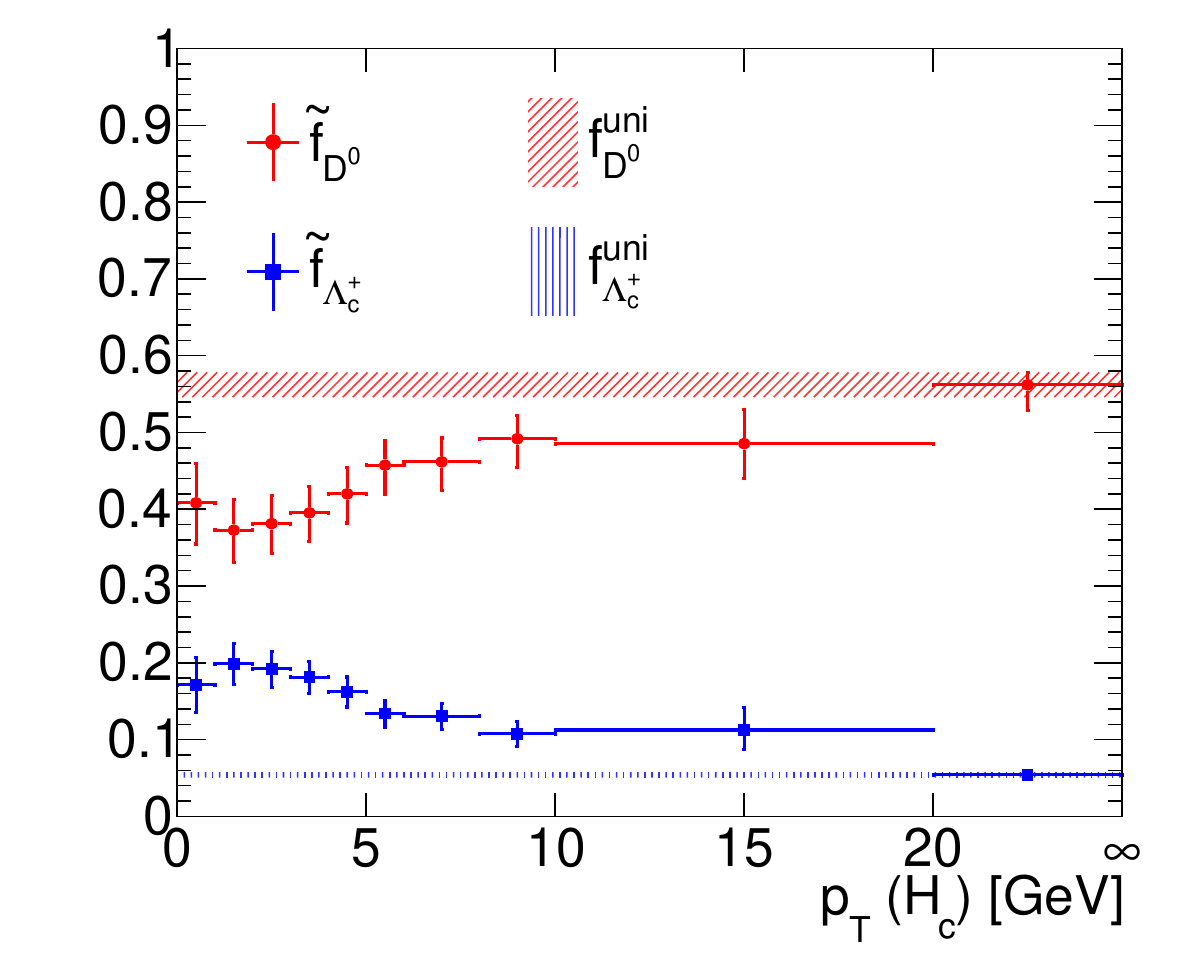}
 \end{center}
 \caption{$\Lambda_c^+/D^0$ measurements from ALICE (left) and CMS (middle), with figures adapted from \cite{ALICE_LcToD0_5TeV_update} and \cite{CMS_Lc_5TeV_update}, respectively. As a reference, the fragmentation fraction of $e^+e^-$ data \cite{fragfrac_comb} was added as the purple band. These measurements were used to derive $p_T$-dependent $D^0$ and $\Lambda_c^+$ production fractions for $pp$ collisions (right).}
\label{fig:ratio_MsToBy_LHCandLEP}
\end{figure}

However, recent reports from LHC experiments, especially ALICE, show large differences for the fragmentation fractions between $pp$ and $e^+e^-$/$ep$ collisions \cite{ALICE_cFragFrac_5TeV}. In particular, a much larger $\Lambda_c^+$ fragmentation fraction is observed in $pp$ data with $\sim 5\sigma$ difference compared to $e^+e^-/ep$ data, while the meson fractions are correspondingly smaller. This is strongly related to a clear $p_T$-dependence of the cross-section ratio $\Lambda_c^+$ \cite{ALICE_LcToD0_5TeV_update, CMS_Lc_5TeV_update} (and $\Xi_c^0$ \cite{ALICE_Xc_5TeV}) to $D^0$ observed in $pp$ collisions in the lower $p_T$ region, while it is asymptotically approaching the $e^+e^-$ data at high $p_T$. This is shown for 5 TeV data in the left and middle figure of Fig.\ref{fig:ratio_MsToBy_LHCandLEP}. 
A similar significant $p_T$-dependence of the baryon-to-meson ratio was also reported from beauty production \cite{LbToB0}. In contrast, no such significant kinematic dependence was observed for meson-to-meson and baryon-to-baryon ratios for charm within the current precision (e.g., \cite{ALICE_DmesonRatios_5TeV}).

\section{\texorpdfstring{$p_T$}{pT}-dependent charm hadron production fractions}

\begin{figure}
 \begin{center}
  \includegraphics[width=0.47\textwidth]{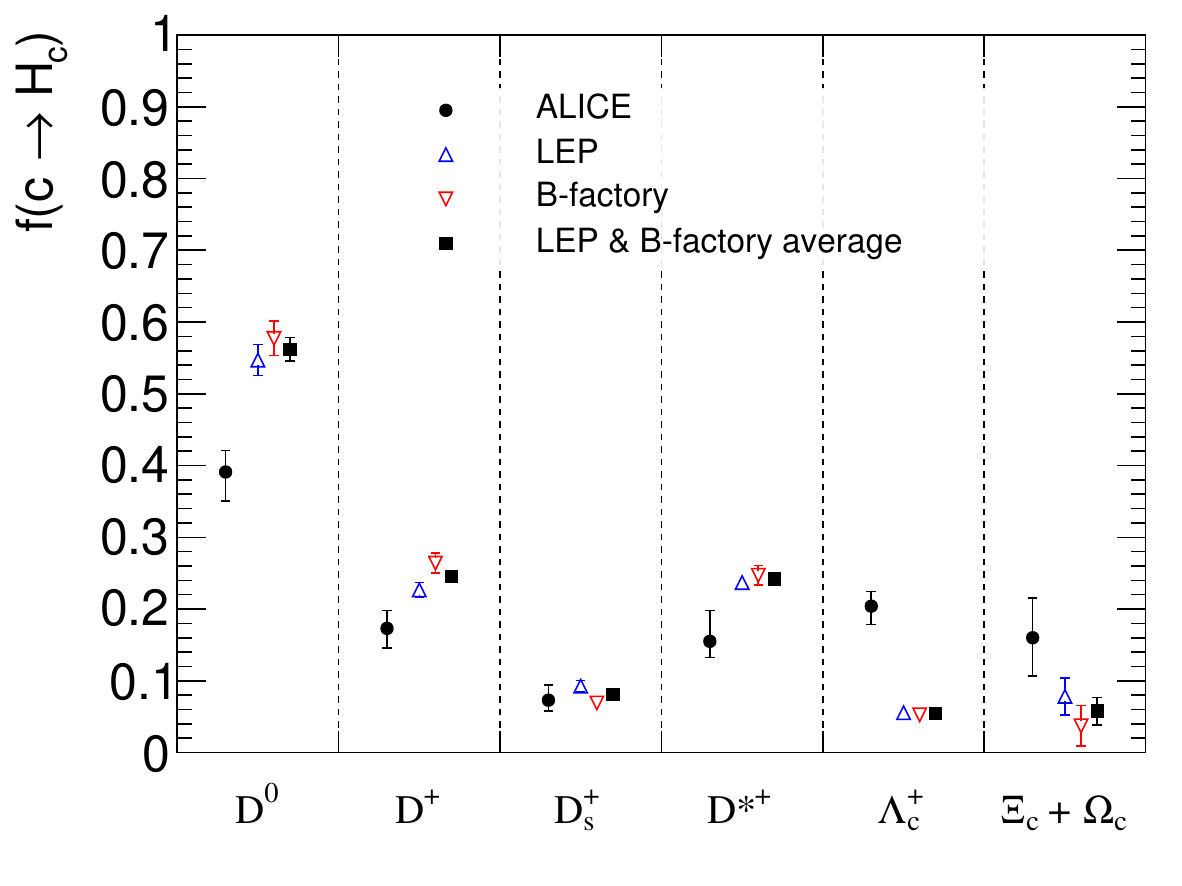}
  \includegraphics[width=0.47\textwidth]{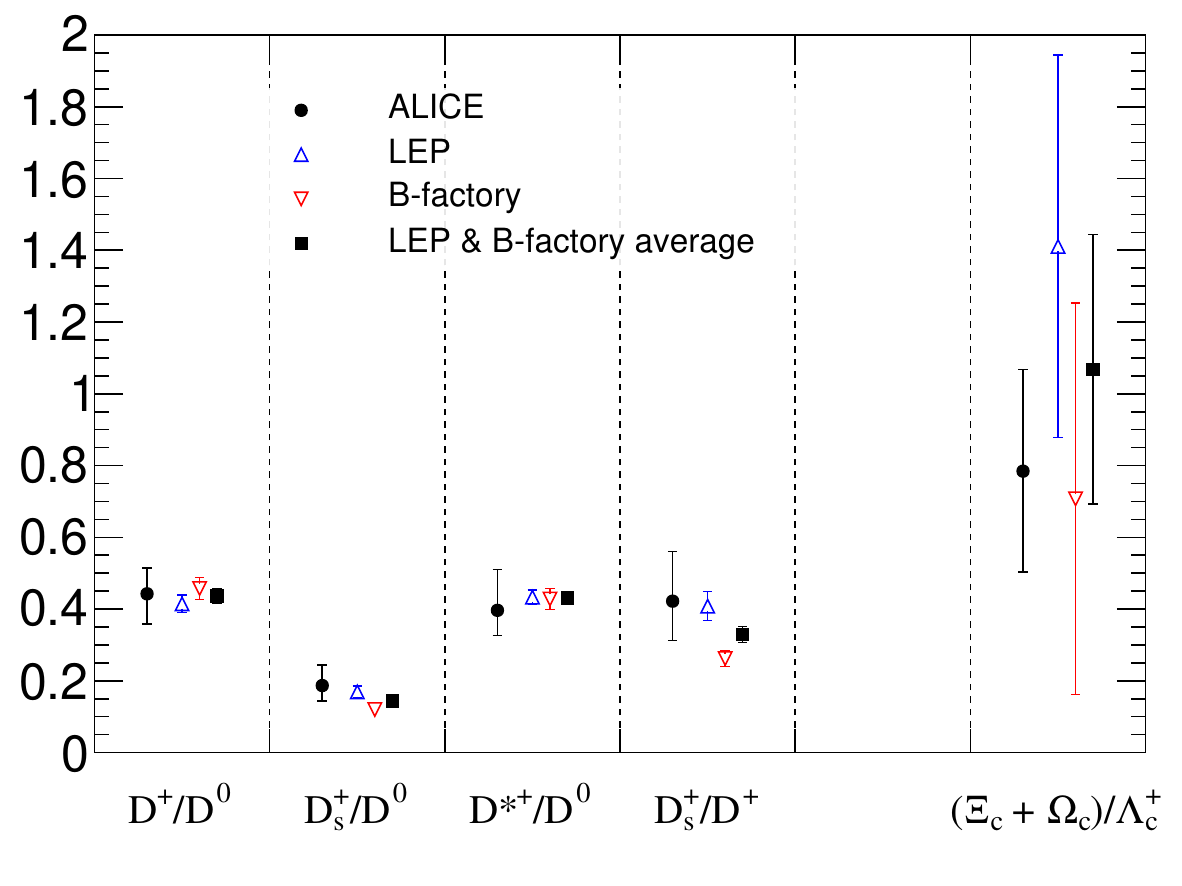}
 \end{center}
 \caption{Charm fragmentation fractions are compared between $pp$ (ALICE) and $e^+e^-$ (LEP and B-factory) collisions in the left figure. The $e^+e^-$ fragmentation fractions of $D^0$, $D^+$, $D_s^+$, $\Lambda_c^+$ are extracted directly from \cite{fragfrac_comb}. The summed fraction of $\Xi_c^0$, $\Xi_c^+$ and $\Omega_c^0$ ($\Xi_c + \Omega_c$) which are not explicitly measured yet from $e^+e^-/ep$ collisions, is derived assuming the sum of all the weakly decaying ground state fractions to be unity. From this, the meson-to-meson and baryon-to-baryon fragmentation fraction ratios are computed and compared in the right figure.} \label{fig:MsToBy}
\end{figure}

Based on these observations, we make some simplifying assumptions consistent with experimental measurements within their uncertainties\footnote{The corresponding uncertainties will all be included in the evaluation of the systematics.}, which are then applied to the extrapolation for $pp$ data. The first assumption is that the meson-to-meson and baryon-to-baryon ratios do not strongly depend on either collision system (see Fig.\ref{fig:MsToBy}) or kinematic range \cite{ALICE_DmesonRatios_5TeV}. 
And the second assumption is that the meson-to-baryon ratio, while $p_T$-dependent (see Fig.\ref{fig:ratio_MsToBy_LHCandLEP}), is not strongly $y$-dependent, which will be indirectly confirmed from the data later in Fig.\ref{fig:xSec_y}. 

We introduce a \emph{$p_T$-dependent hadron production fraction} $\tilde{f}(p_T)$ for $pp$ collisions, which is defined to be asymptotically close to $e^+e^-$ fragmentation fractions\footnote{With fragmentation universality (only), hadron production fractions and fragmentation fractions are the same up to small differences in the fragmentation function.} ($f^{uni}$) at high $p_T$, i.e.,  
\begin{equation} \label{eq:def_ftilde}
 \tilde{f}_{H_c}(p_T) \equiv \frac{d\sigma_{H_c}}{\Sigma_{wd} d\sigma_{H_c}} \equiv f_{H_c}^{uni} F(p_T),
\end{equation}
where $d\sigma_{H_c}$ is now the $p_T$-differential cross-section of a charm hadron state ($H_c$), and the weakly decaying ground states ($wd$) are known to be $D^0$, $D^+$, $D_s^+$, $\Lambda_c^+$, $\Xi_c^0$, $\Xi_c^+$ and $\Omega_c^0$. Here the $p_T$-dependent factor, $F(p_T)$ is constrained to be unity at high $p_T$, and according to the first assumption above, the same $F(p_T)$ is applied to each meson ($F_{MS}(p_T)$) and baryon ($F_{BY}(p_T)$) state. As we observed from the ALICE measurements \cite{ALICE_cFragFrac_5TeV} that meson fractions are smaller respective to $e^+e^-$ collisions while baryon ($\Lambda_c^+$) fraction is bigger, $F_{MS}(p_T)$ is less than 1 at lower $p_T$ and $F_{BY}(p_T)$ is larger than 1. These $F_{MS}(p_T)$ and $F_{BY}(p_T)$ can be expressed by a meson-to-baryon ratio measurement as a function of $p_T$, which are represented by the most precise measurement of $\Lambda_c^+$/$D^0$ from ALICE \cite{ALICE_LcToD0_5TeV_update}, CMS \cite{CMS_Lc_5TeV_update}, and the $e^+e^-$ data \cite{fragfrac_comb} (left and middle figures of Fig.\ref{fig:ratio_MsToBy_LHCandLEP}), and $f_{H_c}^{uni}$s which are shown in Fig.\ref{fig:MsToBy}. As a result, e.g., $\tilde{f}_{D^0}(p_T)$ and $\tilde{f}_{\Lambda_c^+}(p_T)$ are derived by Eq.(\ref{eq:def_ftilde}), and shown in the right figure of Fig.\ref{fig:ratio_MsToBy_LHCandLEP}. For the $\tilde{f}$ uncertainties, an additional uncertainty is assigned to account for a possibly $p_T$-dependent ratio of $D_s^+$ to the other mesons. This is motivated by what was observed with much better precision in B production measurements which show a moderate but clear $p_T$-dependence of the ratio of $B_s^0$ to the other B mesons \cite{LHCb_Bmesons, CMS_Bmesons_13TeV}.

\section{Extrapolation for \texorpdfstring{$pp$}{pp} collisions with non-universal charm fragmentation}

For the total charm cross section, charm hadron measurements in a constrained kinematic range need to be extrapolated and/or interpolated to the full kinematic range. As an example for this procedure, the $D^0$ measurements at 5 TeV from ALICE \cite{ALICE_Dmesons_5TeV} and LHCb \cite{LHCb_Dmesons_5TeV} are extrapolated using FONLL theory. To account for fragmentation non-universality, the original FONLL calculation, which is based on the assumption of charm fragmentation universality ($d\sigma_{H_c}^{\text{FONLL}}$), is modified by applying the $p_T$-dependent production fraction, $\tilde{f}_{H_c}(p_T)$ of Eq.(\ref{eq:def_ftilde}) ($d\sigma_{H_c}^{\text{FONLL with } \tilde{f}}$):
\begin{equation} \label{eq:modFonll}
 d\sigma_{H_c}^{\text{FONLL}} = f_{H_c}^{uni} \cdot \bigg(d\sigma_{pp \rightarrow c\bar{c}}^{\text{FONLL}} \otimes D_{c \rightarrow H_c}^{\text{NP}}\bigg), \quad d\sigma_{H_c}^{\text{FONLL with } \tilde{f}} = \tilde{f}_{H_c}(p_T) \cdot \bigg(d\sigma_{pp \rightarrow c\bar{c}}^{\text{FONLL}} \otimes D_{c \rightarrow H_c}^{\text{NP}}\bigg),
\end{equation}
where the charm quark fragmentation function distribution $D_{c \rightarrow H_c}^{\text{NP}}$ is normalized to unity. Note that the data driven $p_T$ dependence of $\tilde{f}_{H_c}$ blurs the physical meaning of the parameters entering $d\sigma$ and $D^{\text{NP}}$, even though their technical definition remains the same. Since the calculation thereby loses the character of a QCD prediction, and reduces to a theory-inspired parametrization of the cross section for the sole purpose of extrapolation and interpolation, {\sl all} its parameters should and can now be determined in a data driven way.

Redoing a fit of the PDFs is outside of the scope of this work. Since the FONLL calculation uses a variable flavour number scheme (VFNS), the ideal PDF for this work would be the VFNS version of the PROSA PDF \cite{prosa2015, prosa2019}, which includes a fit to ALICE and LHCb charm data in a way that is {\sl not} affected by the $p_T$ dependence/non-universality of charm fragmentation as used in this work. Unfortunately, only the central value for this PDF is available, while uncertainties are available only for the fixed-flavour version. Fortunately, it turns out that the older CTEQ6.6 PDF \cite{cteq66} happens to be consistent with the PROSA PDF for both central value and uncertainty. We thus pragmatically use this PDF as a proxy for the PROSA\_VFNS PDF. 

\begin{figure}
 \begin{center} 
  \includegraphics[width=0.24\textwidth]{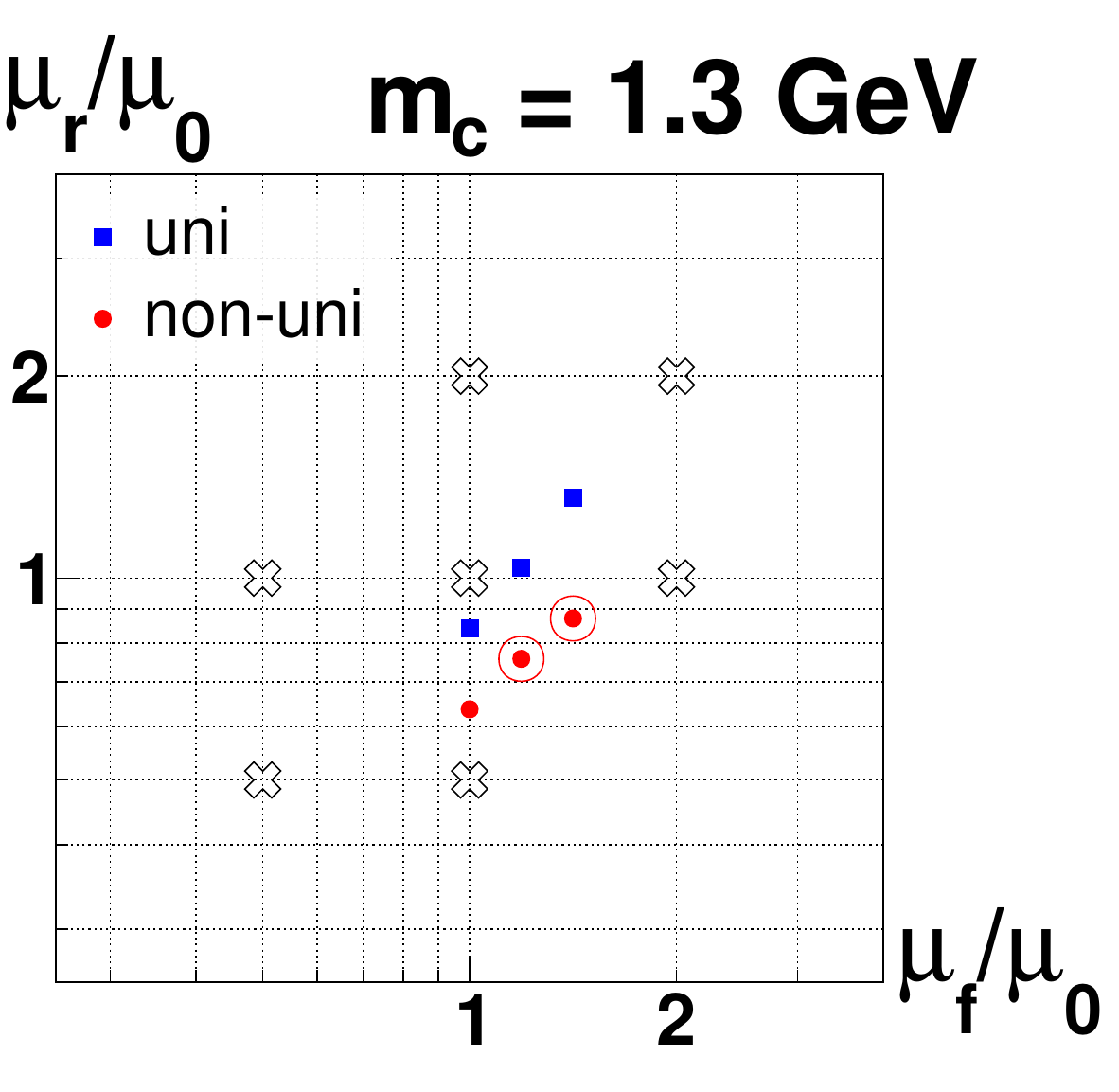}
  \includegraphics[width=0.24\textwidth]{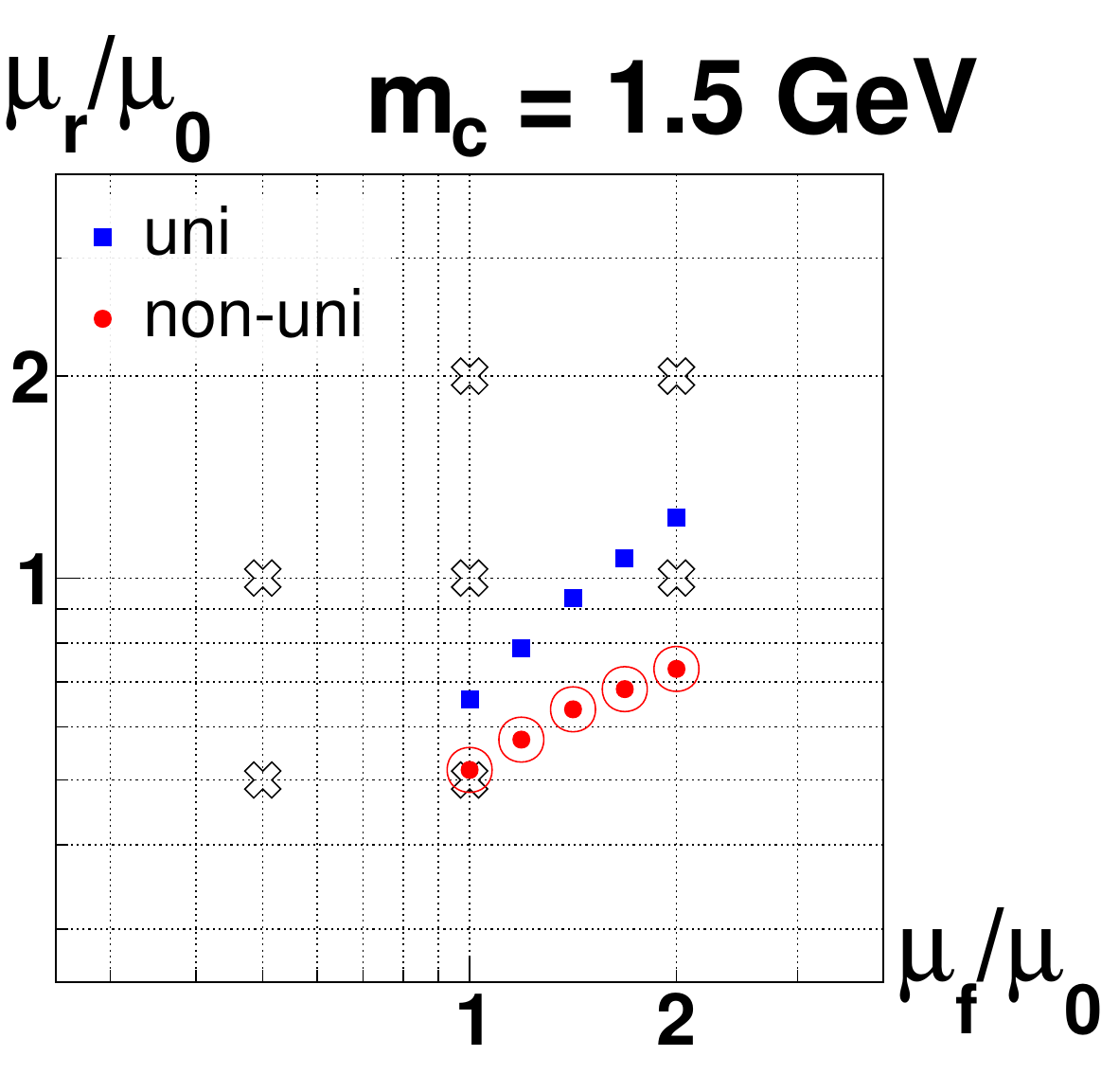}
  \includegraphics[width=0.24\textwidth]{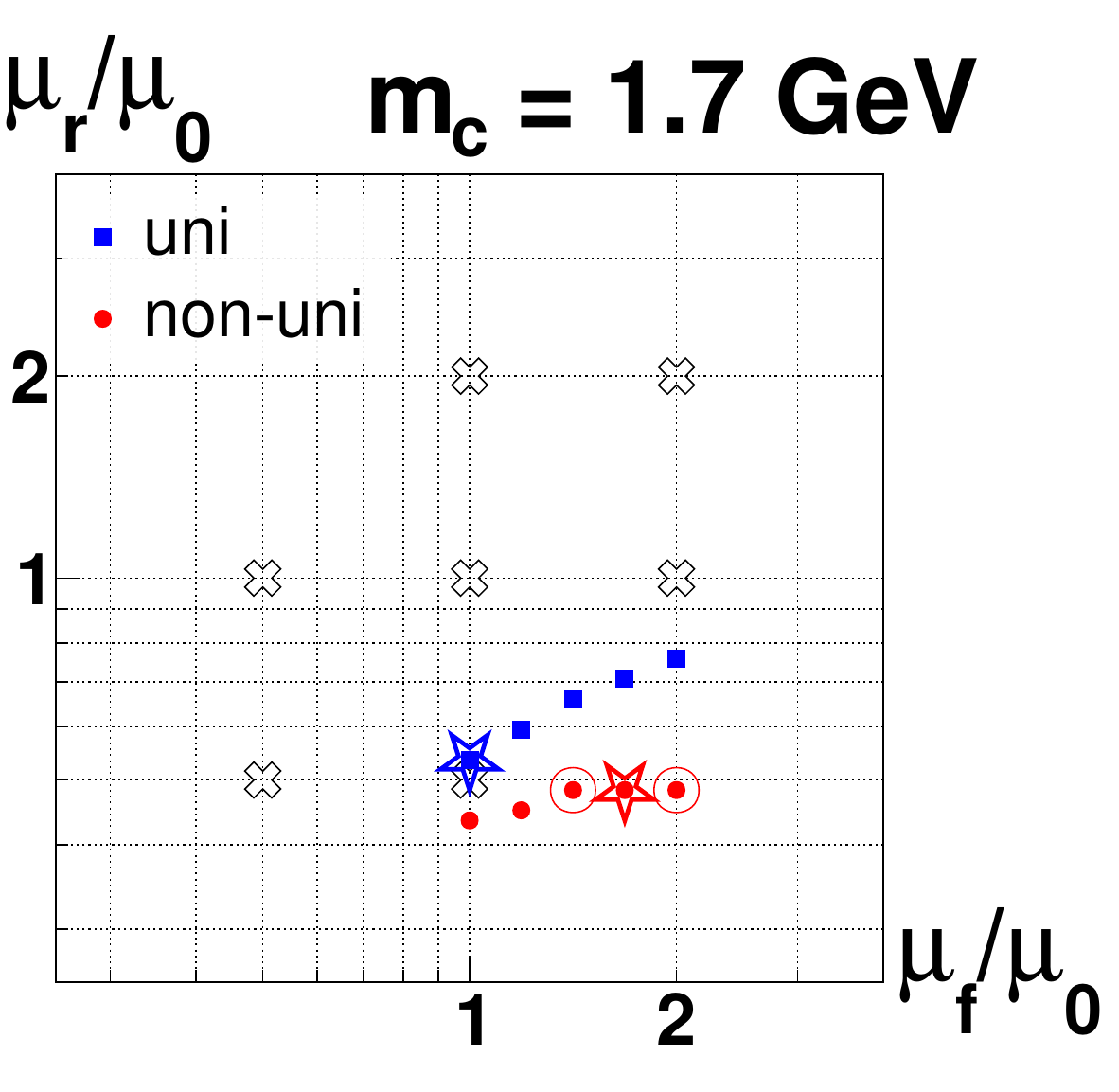}
  \includegraphics[width=0.24\textwidth]{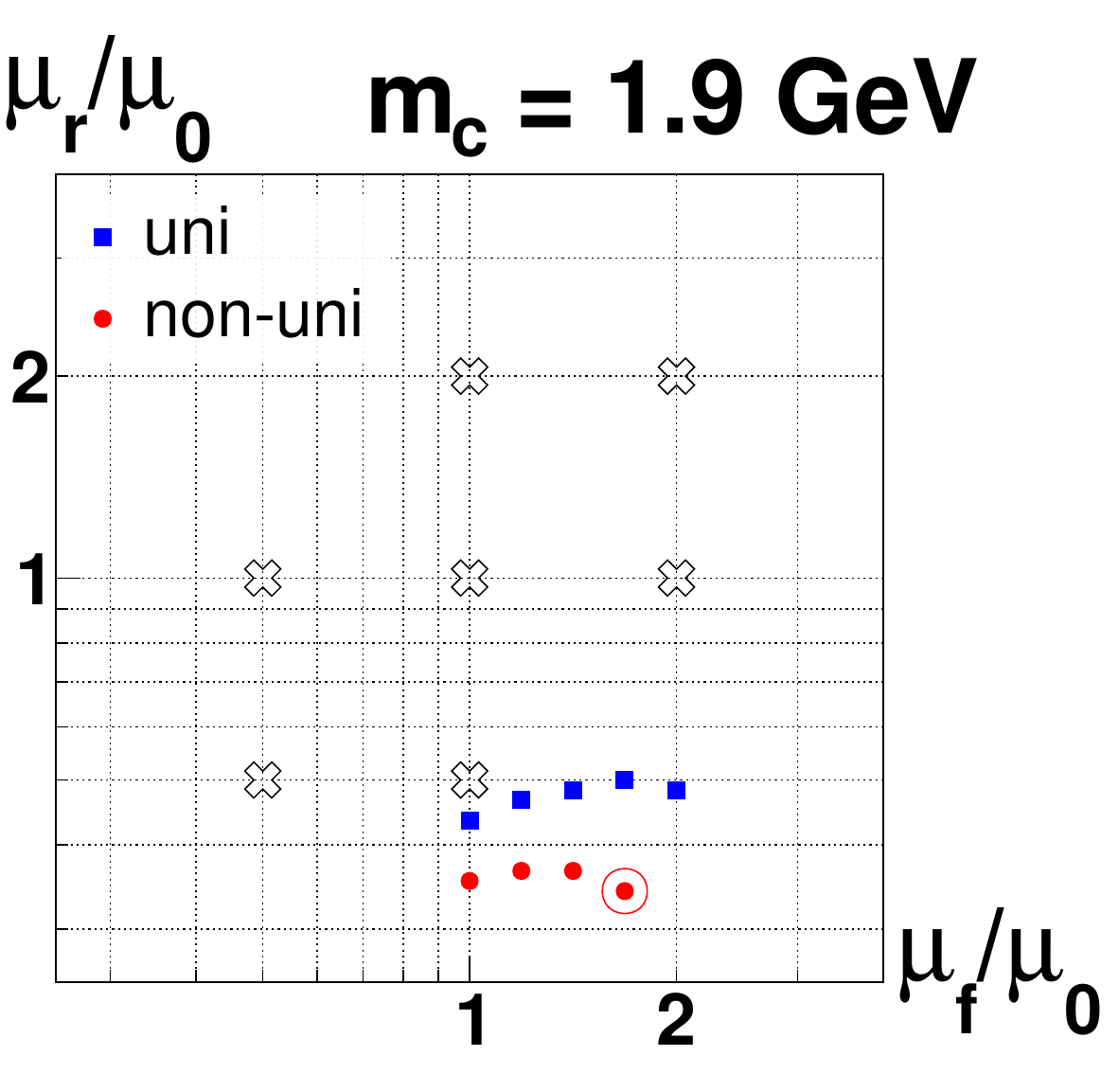}
 \end{center}
 \caption{$\chi^2$ scan results for the $D^0$ measurements at 5 TeV. In addition to the core results including fragmentation non-universality (red circle points), the results from the same procedure with the traditional universality assumption are also shown (blue square points). The respective best parameters are marked by a star. For the non-universality case, the points within the $\chi^2$ scan uncertainty contour are marked by additional red outer circles. The cross marks represent the usual 7-point scale variation and are shown here for comparison only.} \label{fig:scales_chosen}
\end{figure}

\begin{figure}
 \begin{center}
 \includegraphics[width=0.16\textwidth]{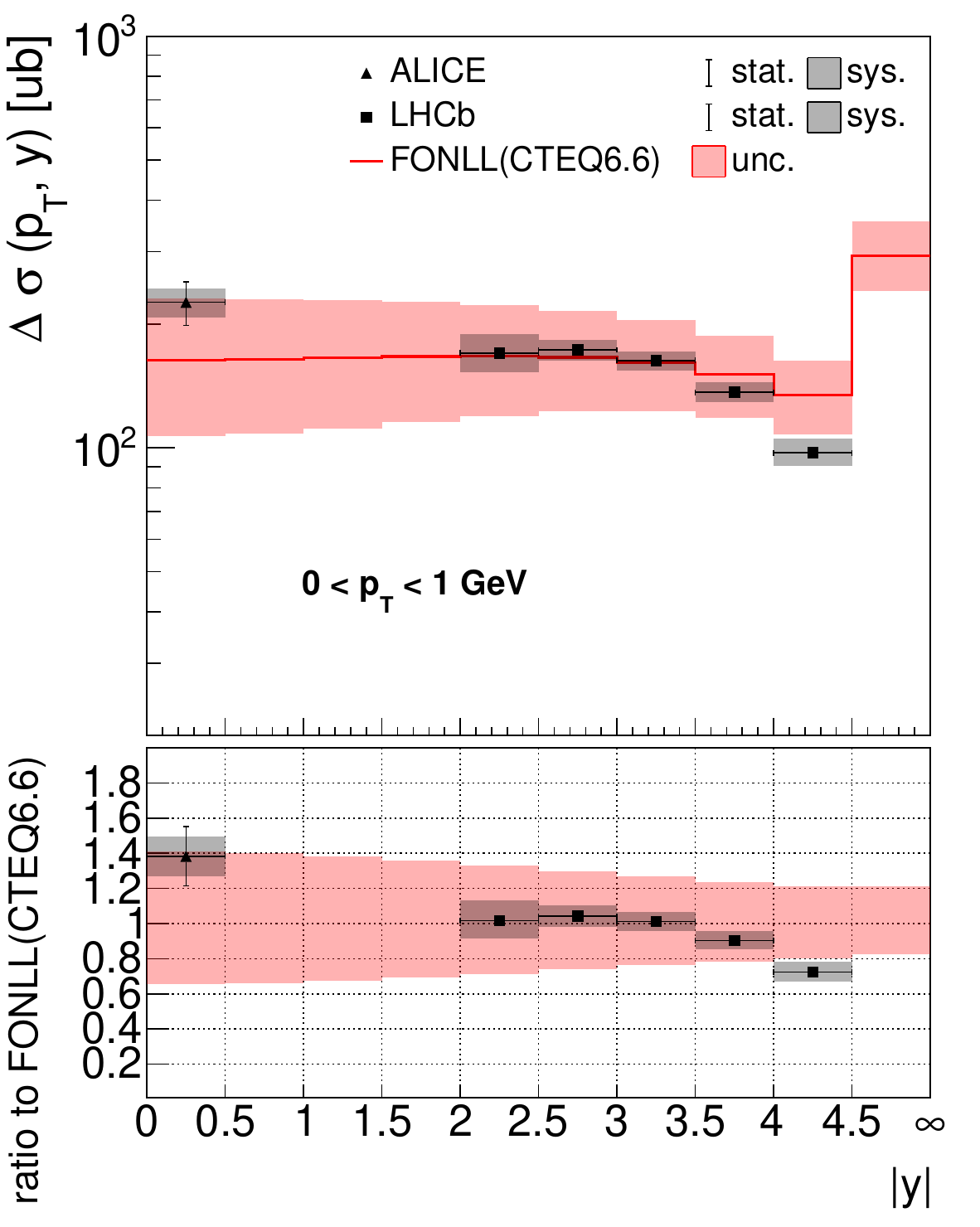}
 \includegraphics[width=0.16\textwidth]{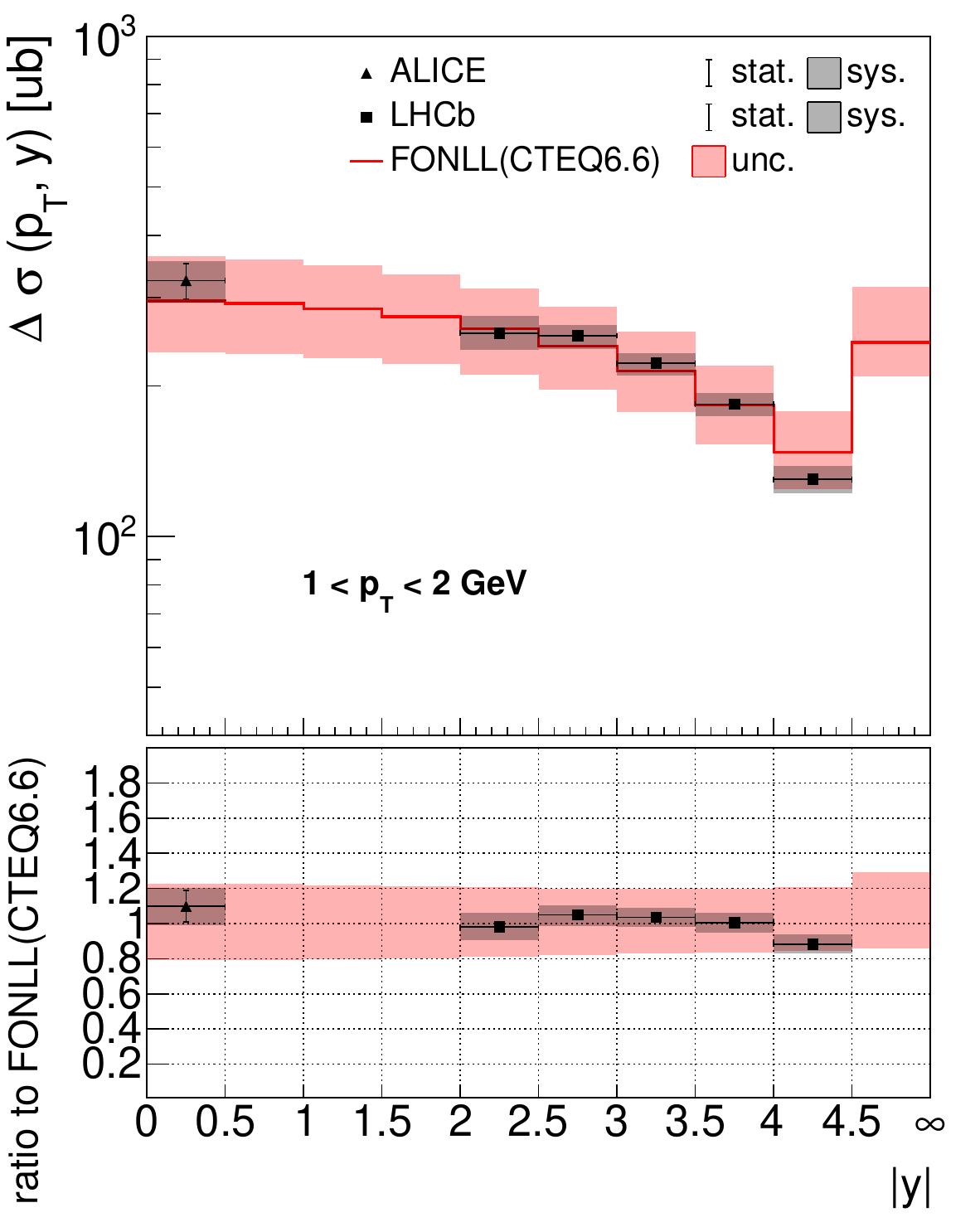}
 \includegraphics[width=0.16\textwidth]{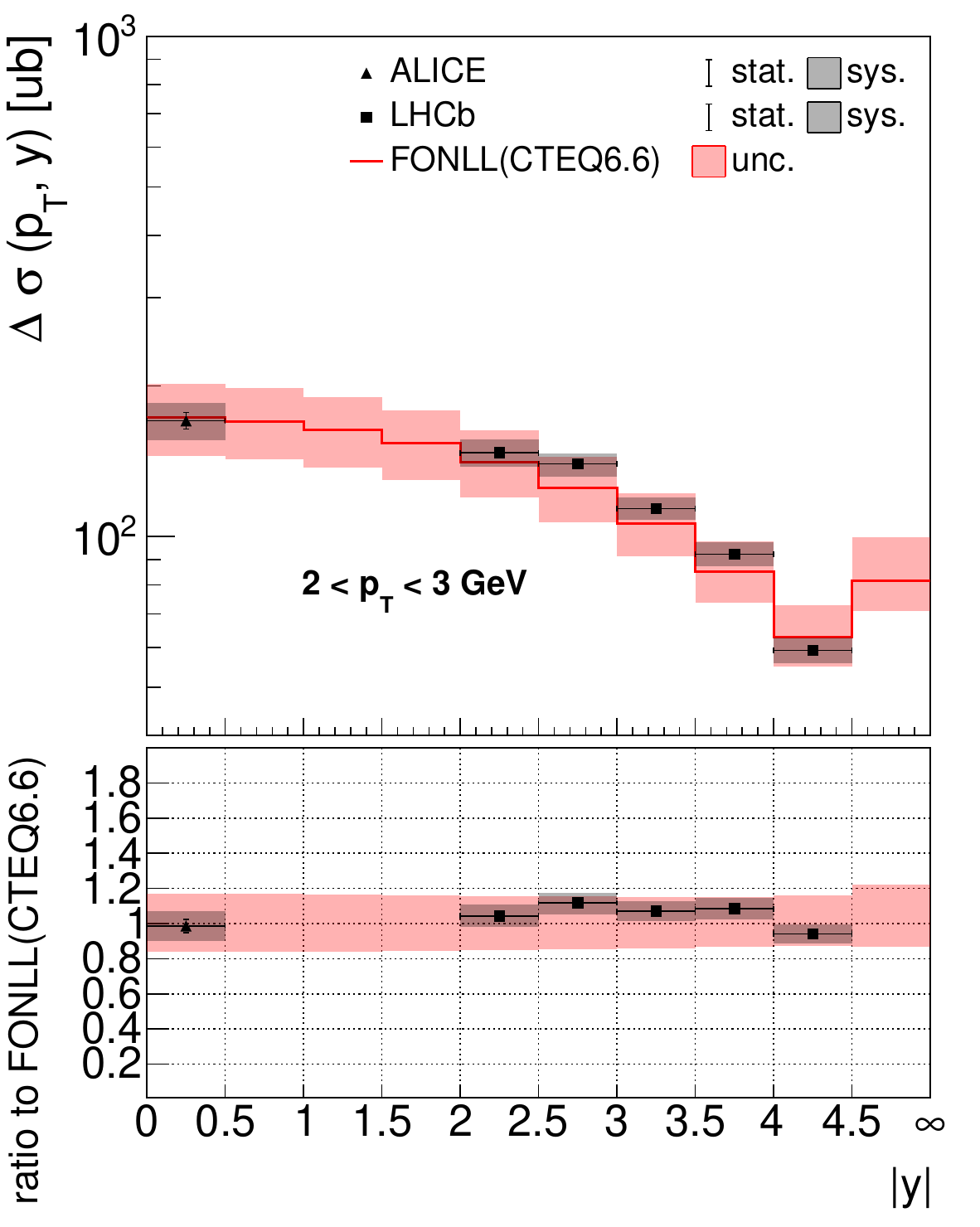}
 \includegraphics[width=0.16\textwidth]{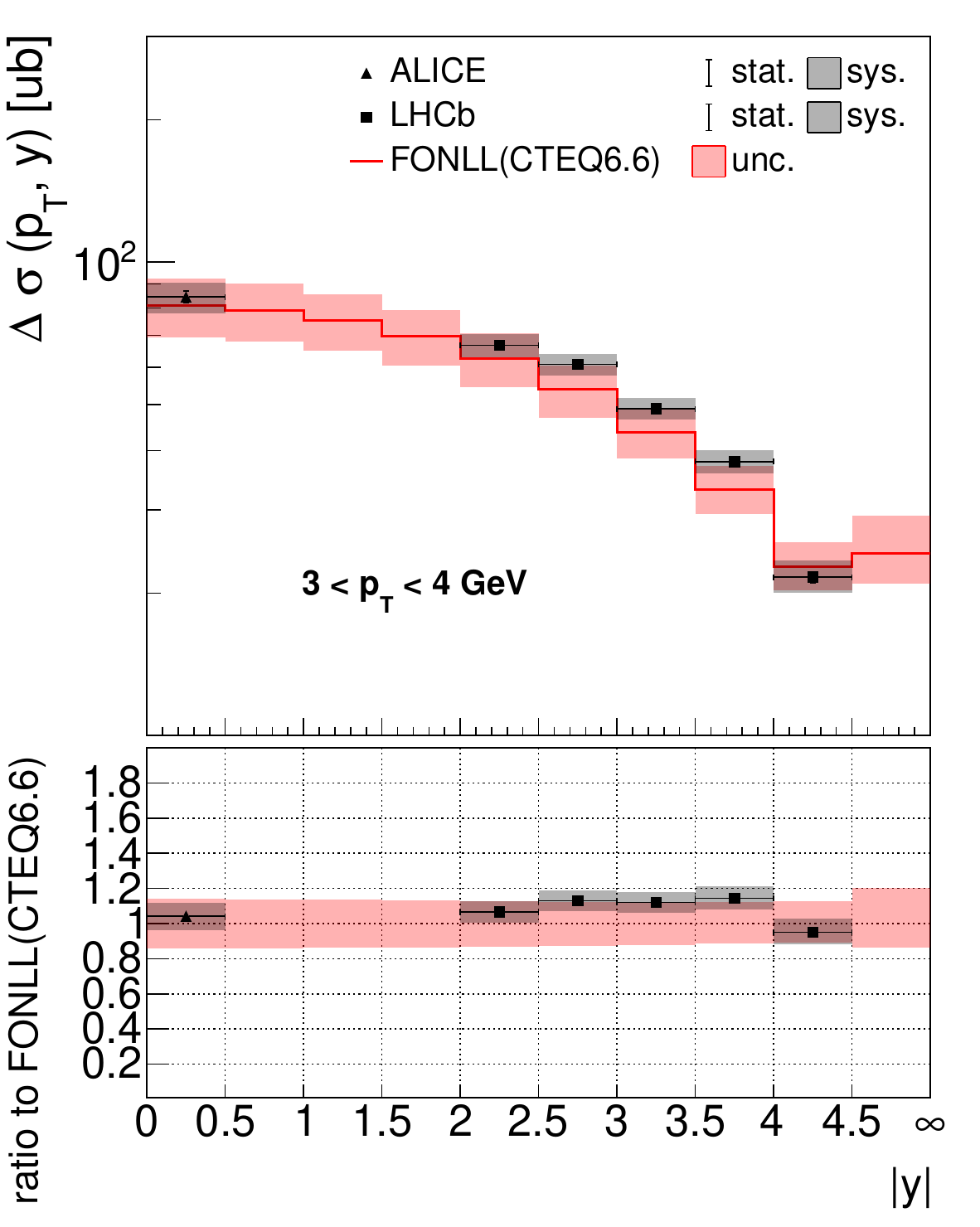}
 \includegraphics[width=0.16\textwidth]{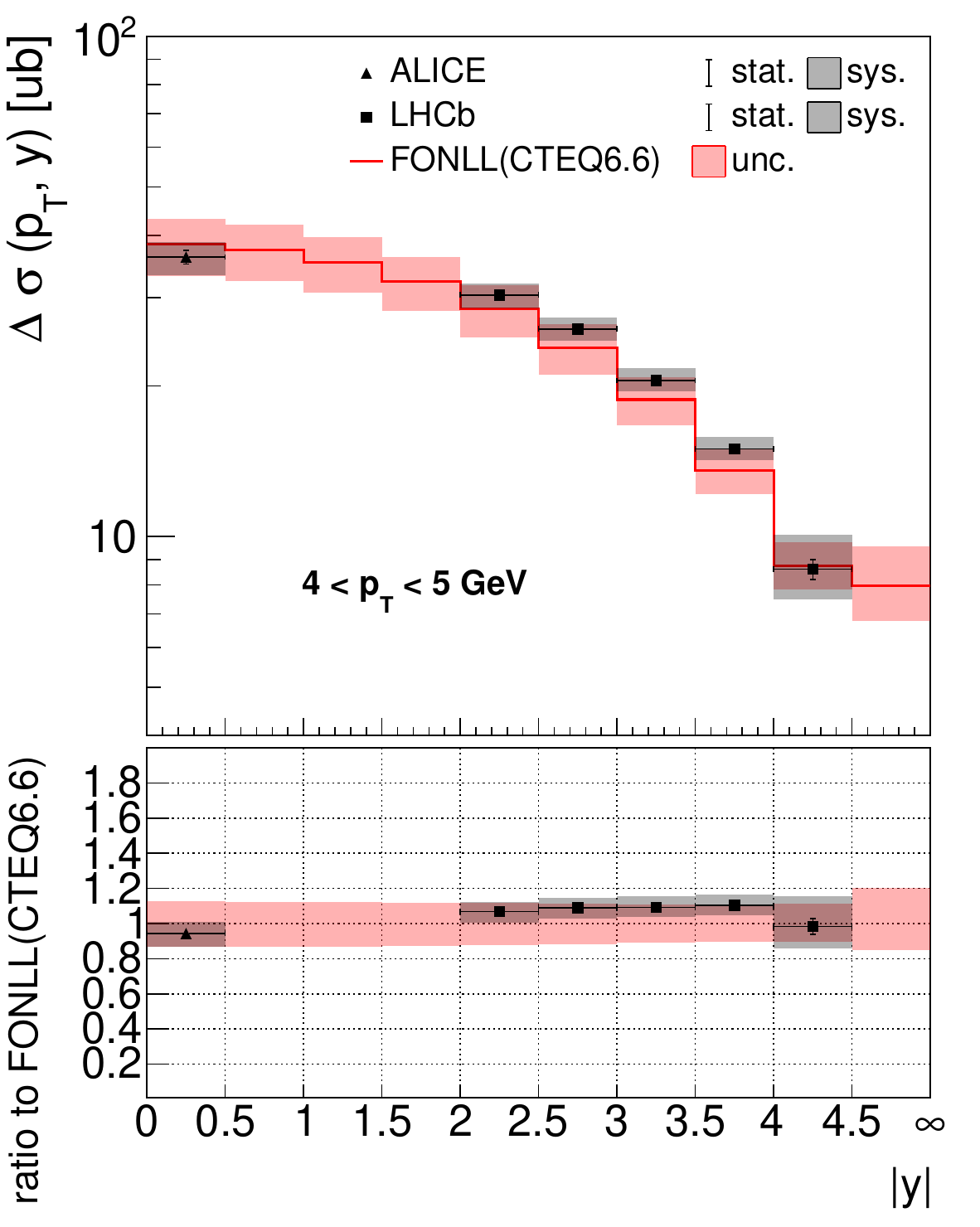}
 \includegraphics[width=0.16\textwidth]{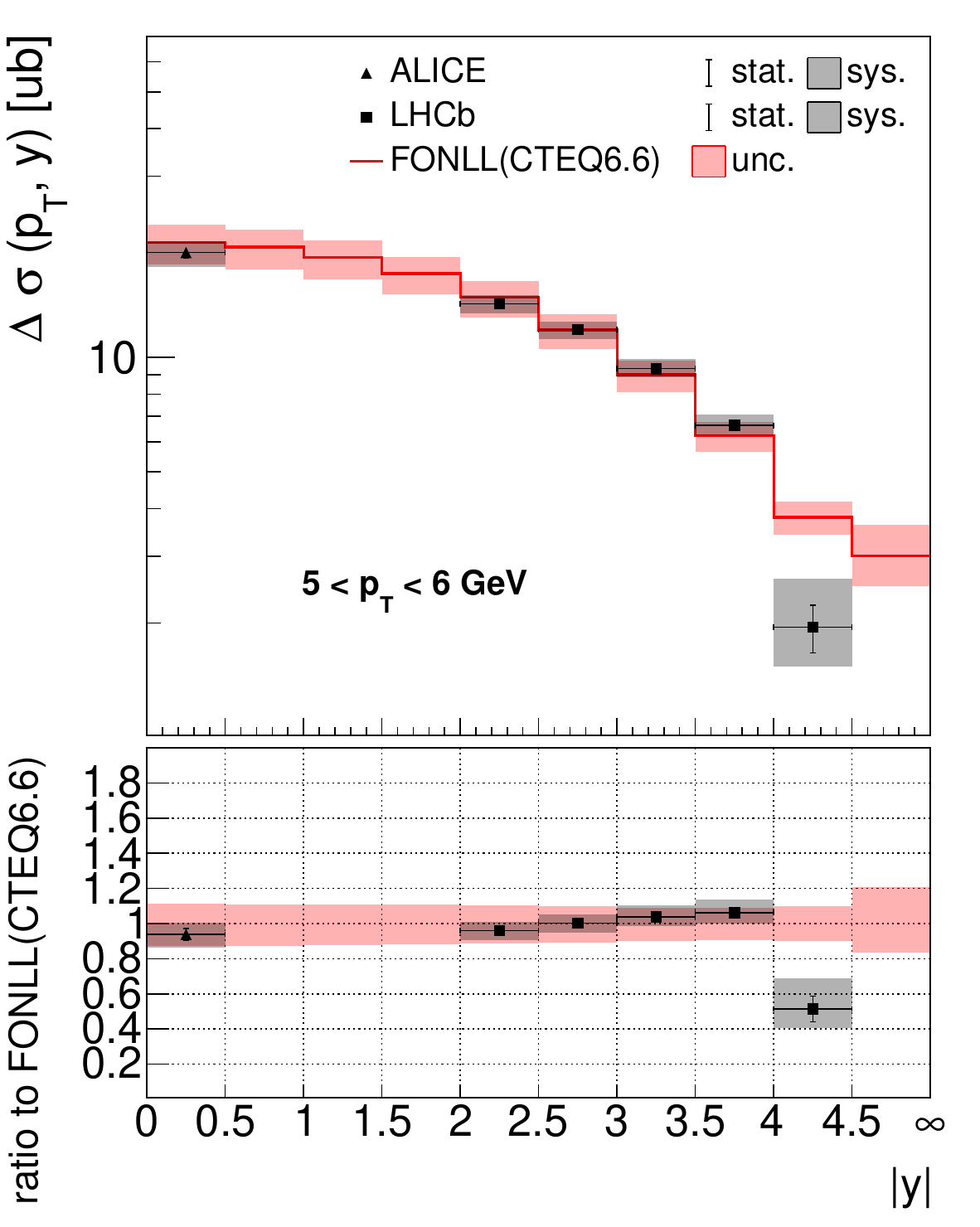}
 \includegraphics[width=0.16\textwidth]{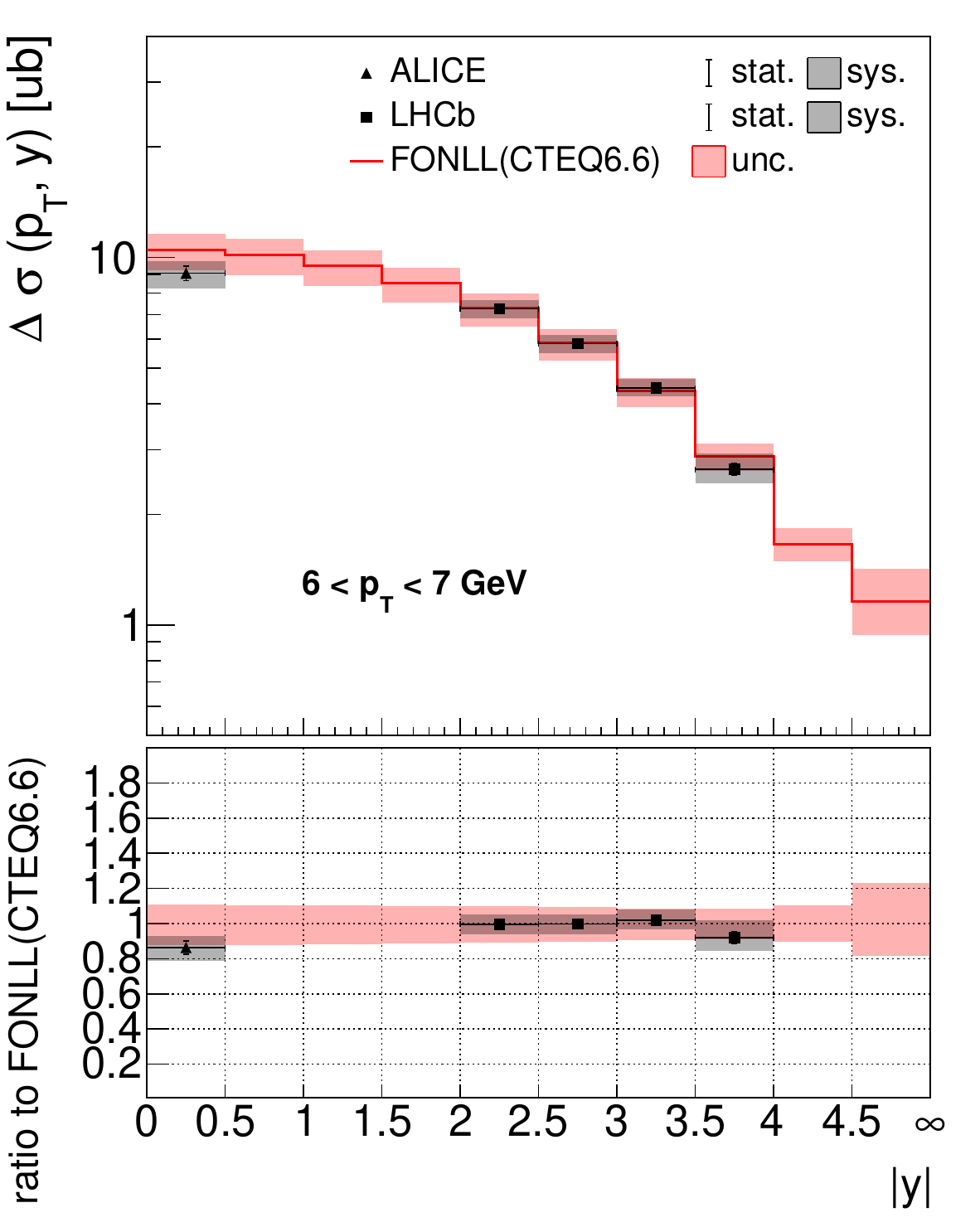}
 \includegraphics[width=0.16\textwidth]{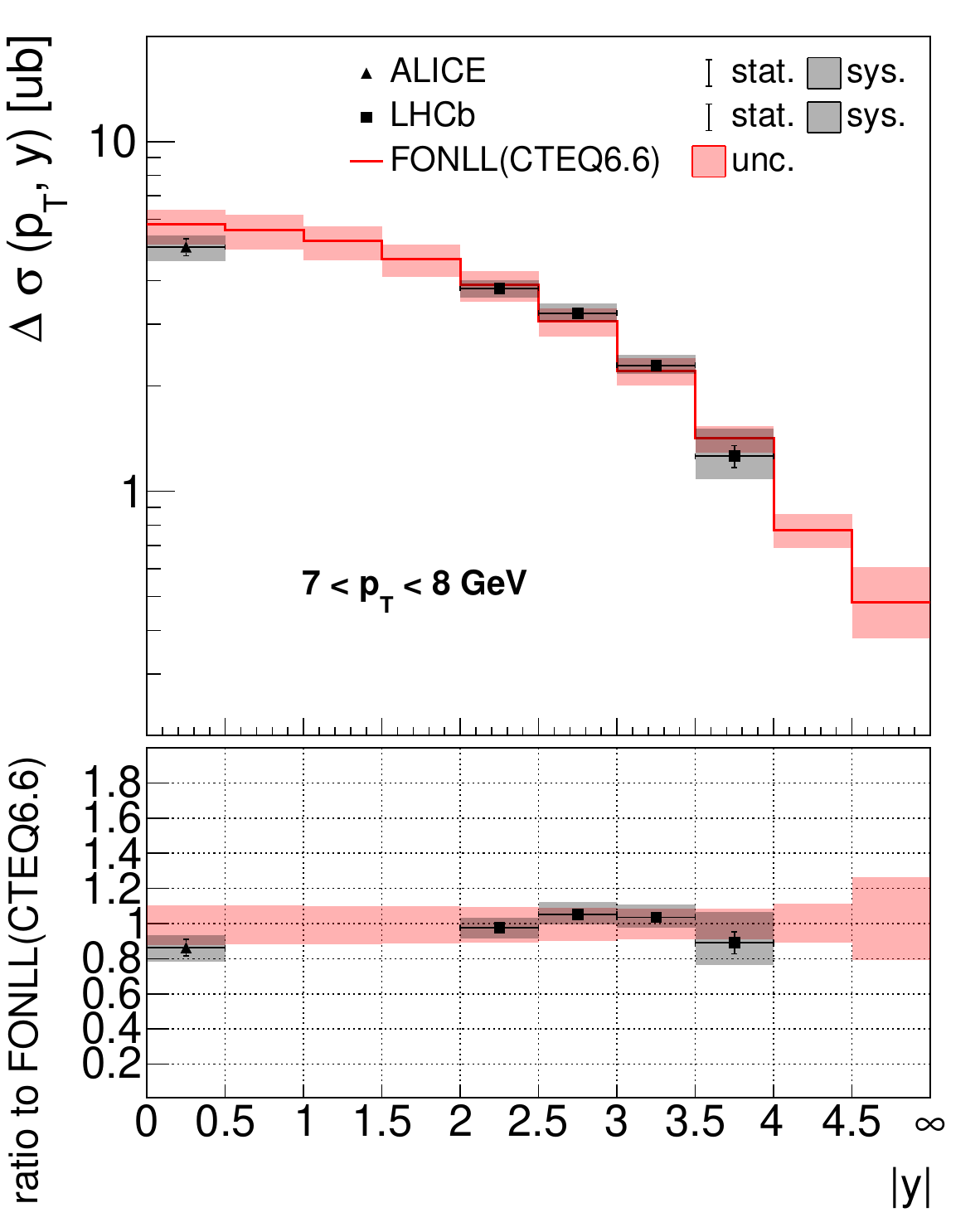}
 \includegraphics[width=0.16\textwidth]{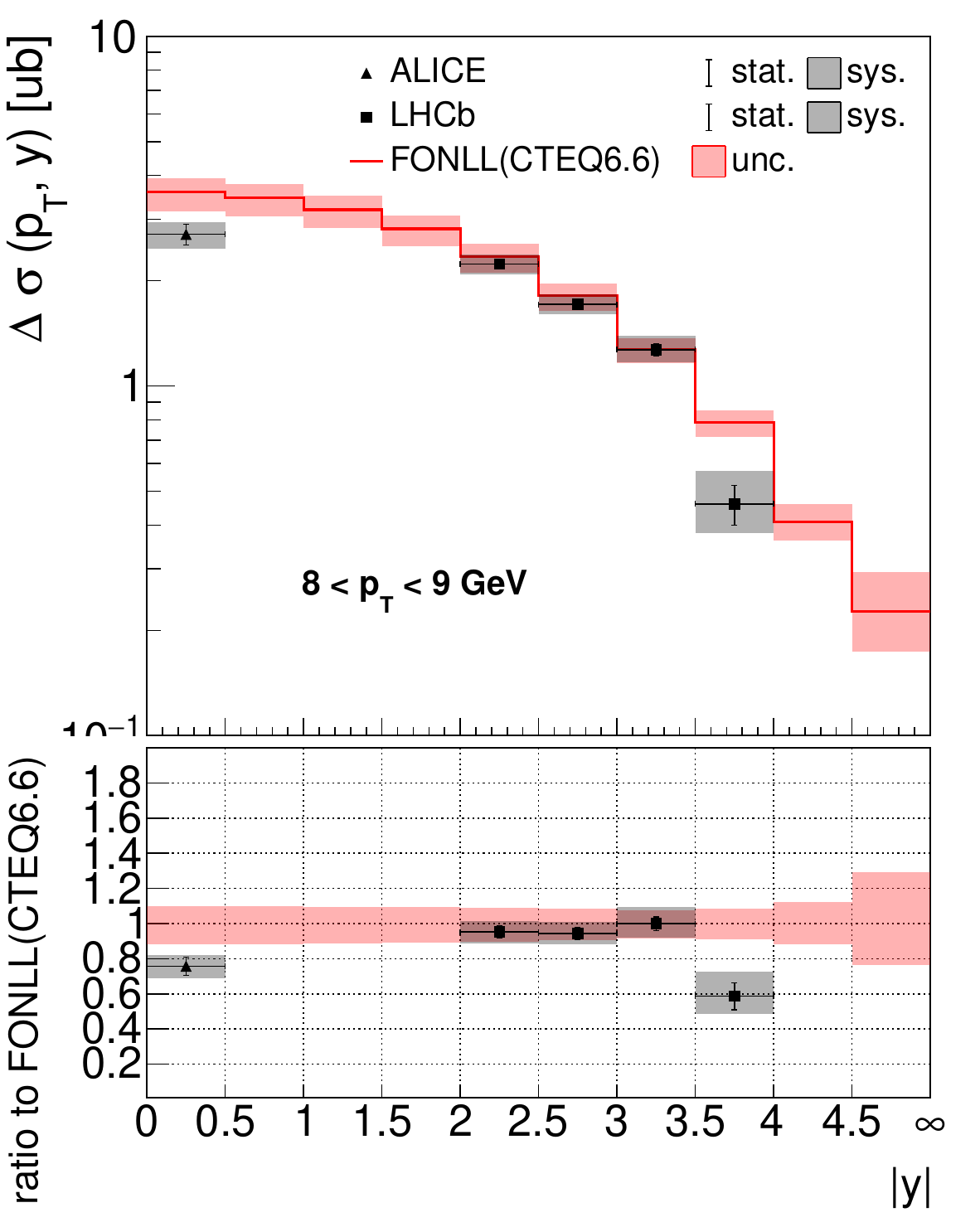}
 \includegraphics[width=0.16\textwidth]{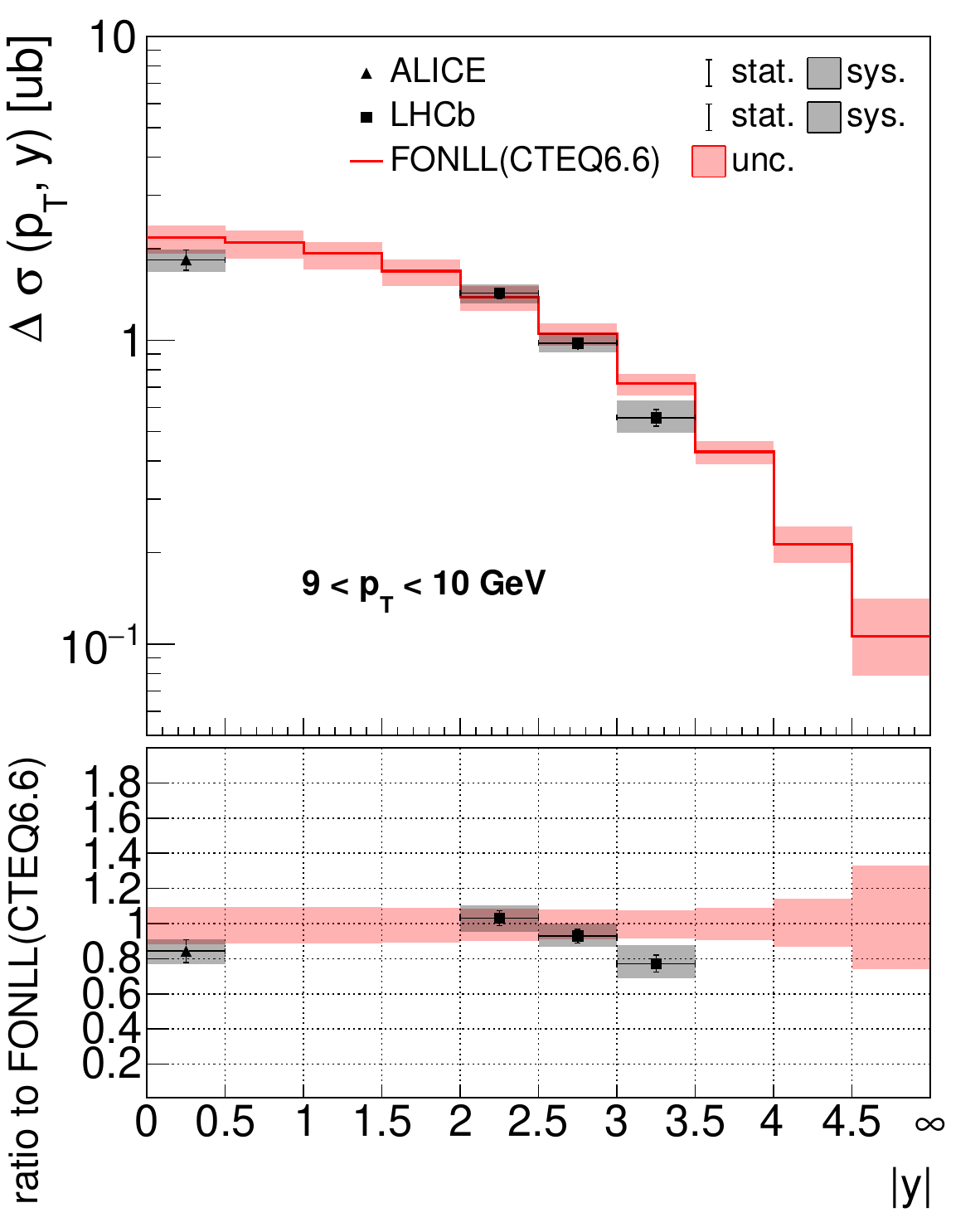}
 \includegraphics[width=0.16\textwidth]{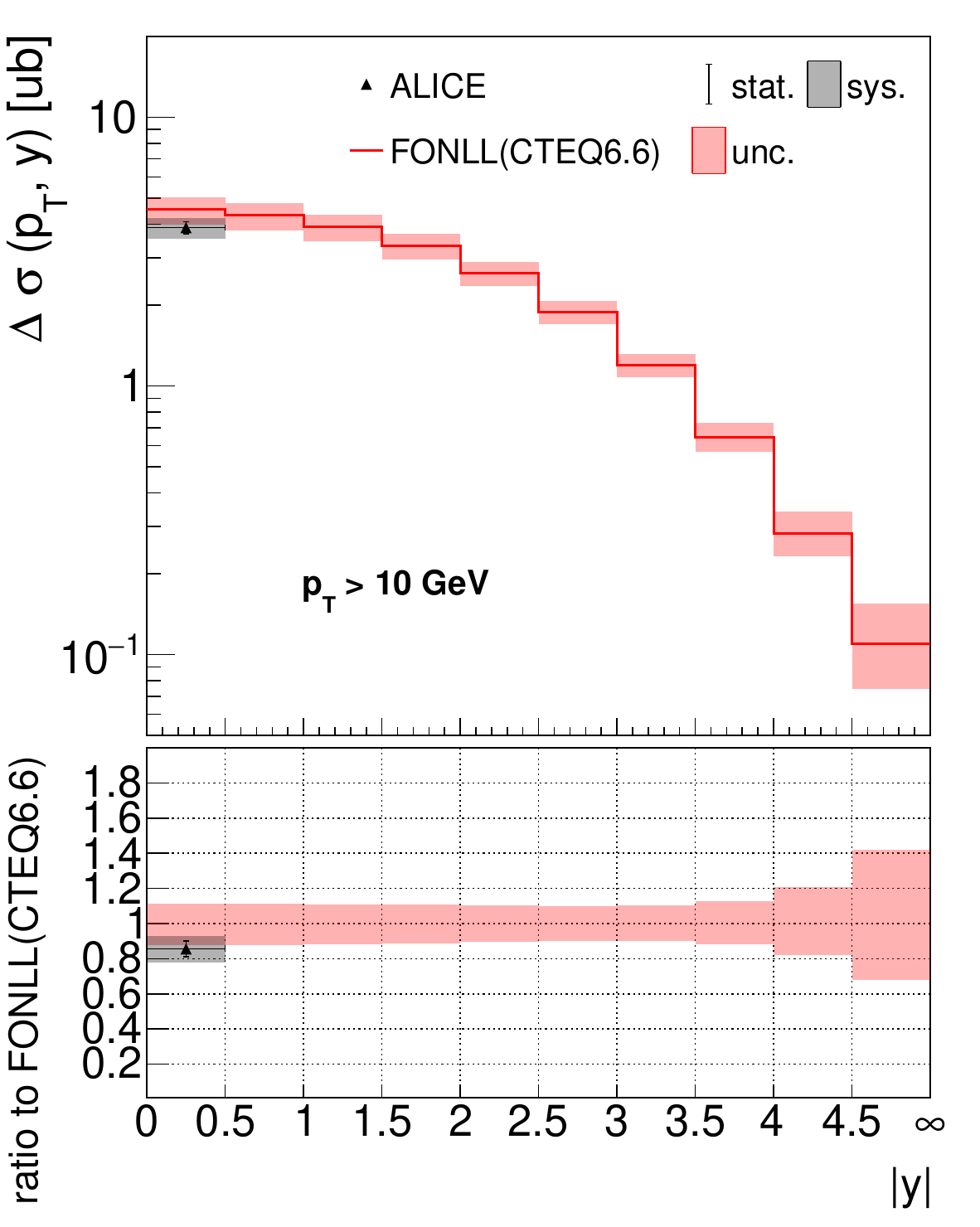}
 \end{center}
 \caption{$D^0+\overline{D}{}^0$ cross-sections as a function $|y|$ in bins of $p_T$. The red bands are the data driven FONLL as obtained with non-universal charm fragmentation, which describe the data (black points/grey boxes) well in the full phase space. The total uncertainty of the data driven FONLL includes the uncertainties of the CTEQ6.6 PDF \cite{cteq66}, $\tilde f$ (Fig. \ref{fig:ratio_MsToBy_LHCandLEP}), and the $\chi^2$ scan (Fig. \ref{fig:scales_chosen}).} \label{fig:xSec_y}
\end{figure}

We parametrize $D_{c\rightarrow H_c}^{\text{NP}}$ by the Kartvelishvili function \cite{fragfunc_kart} with a single parameter\footnote{This parameter is taken to be independent on kinematic in this work.}, $\alpha_K$. The core FONLL perturbative QCD calculation has three parameters: the factorization scale ($\mu_f$), the renormalization scale ($\mu_r$), and the charm pole mass ($m_c$). To find the best simultaneous description of the ALICE and LHCb double differential $D^0$ data, a $\chi^2$ scan (fit) of these four free parameters is performed, with the PDFs initially fixed to their central value. The best parameters ($\mu_f^b$, $\mu_r^b$, $m_c^b$ and $\alpha_K^b$) are determined by the least $\chi^2$ of the 4-dimensional scan, and \emph{data driven FONLL} is defined as $d\sigma_{H_c}^{\text{FONLL (with } \tilde{f})}(\mu_f^b, \mu_r^b, m_c^b, \alpha_K^b)$. These best parameters and their uncertainty ranges are summarized in Fig.\ref{fig:scales_chosen}, where the least $\chi^2$ results of the 3-dimensional scans with $\mu_f$, $\mu_r$ and $\alpha_K$ for the respective fixed $m_c$ are projected into the 2-dimensional planes ($\mu_f$, $\mu_r$).
In the figure, the best parameter set is marked by a star. The parameter sets which are located within the 4-dimensional 68\% c.l. contour\footnote{With an S-factor of 1.46 to compensate a small deviation from the optimal $\chi^2$/ndof} determined by $\Delta \chi^2 \sim 1\sigma$ are marked by additional outer circles. The $\alpha_K$ uncertainty, which is not shown in the figure, is $6 \lesssim \alpha_K \lesssim 25$. These uncertainties reasonably cover the conventional choices of the parameters for the FONLL predictions \cite{fonll1, fonll2}, although their meaning is somewhat blurred as explained earlier. $d\sigma_{H_c}^{\text{FONLL with } \tilde{f}}(\mu_f^b, \mu_r^b, m_c^b, \alpha_K^b)$ is integrated over each bin in $p_T$ and $|y|$, including overflow bins, and shown with its total uncertainty by red bands in Fig.\ref{fig:xSec_y}. In the same way, all the available $D^0$ measurements from ALICE and LHCb are collected and shown by triangle and square points, respectively, in the figure. Note the consistency with the `$p_T$ dependence only' assumption within this uncertainty. 

\section{Total charm cross section}

To derive the total charm cross section, the total $D^0+\overline{D}{}^0$ cross section is determined first, in two pieces. For the first piece all the available measurements (grey boxes in Fig.\ref{fig:xSec_y}) are summed up, yielding a fiducial cross section of $3.64^{+0.19}_{-0.19}(\text{data}\footnotemark[5])$ mb. Only for the remaining non-measured bins, the data driven FONLL values (red bands in Fig.\ref{fig:xSec_y}) are summed instead, giving the complementary fiducial cross section of $2.95 ^{+0.31}_{-0.33}(\tilde{f}) ^{+0.52}_{-0.44}(\text{PDF})$ mb. The extrapolation factor is thus about 1.7. The two are then added to obatin the total $D^0+\overline{D}{}^0$ cross section, and divided by the measured $D^0$ fragmentation fraction of $pp$ collisions, $f_{D^0}^{pp} = 0.391^{+0.030}_{-0.041}$\footnotemark[5] (\cite{ALICE_cFragFrac_5TeV}, black circle point in the left figure of Fig.\ref{fig:MsToBy}) with an extra factor 2 to average $D^0$ and $\overline{D}{}^0$ state.\footnotetext[5]{The uncertainties are a quadrature sum of statistical and systematic uncertainty.} Finally, the total charm-quark pair cross section at 5 TeV then turns out to be
\begin{equation}
 \sigma_{c\bar{c}} = 8.43 ^{+0.25}_{-0.25}\text{(data)} ^{+0.40}_{-0.42}(\tilde{f}) ^{+0.67}_{-0.56}\text{(PDF)} ^{+0.13}_{-0.12}(\mu_f, \mu_r, m_c, \alpha_K) ^{+0.65}_{-0.88}(f^{pp}) \text{[mb]}.
\end{equation}
The total uncertainty is determined to be $8.43 ^{+1.05}_{-1.16}(\text{total})$ mb.

As a reference, the total charm cross section is rederived also with the charm fragementation universality assumption but without determining dedicated uncertainties for PDFs and $\chi^2$ scan. The $D^0$ cross section for the non-measured kinematic region turns out to be $2.92 ^{+0.08}_{-0.08}(f^{uni})$ mb with $d\sigma_{H_c}^{\text{FONLL}}$. Then the hypothetical total charm cross section at 5 TeV is determined to be $5.84 ^{+0.17}_{-0.17}\text{(data)} ^{+0.17}_{-0.17}(f^{uni})$ mb by $f_{D^0}^{uni} = 0.562 \pm 0.016$ (black square point in the left figure of Fig.\ref{fig:MsToBy}). This is consistent with the value $5.25^{+0.35}_{-0.26}$ mb obtained in \cite{nnloCharm1}, which was also assuming fragmentation universality, using a slightly different set of extrapolation parameters. 

\begin{figure}
 \begin{center}
  \includegraphics[width=0.5\textwidth]{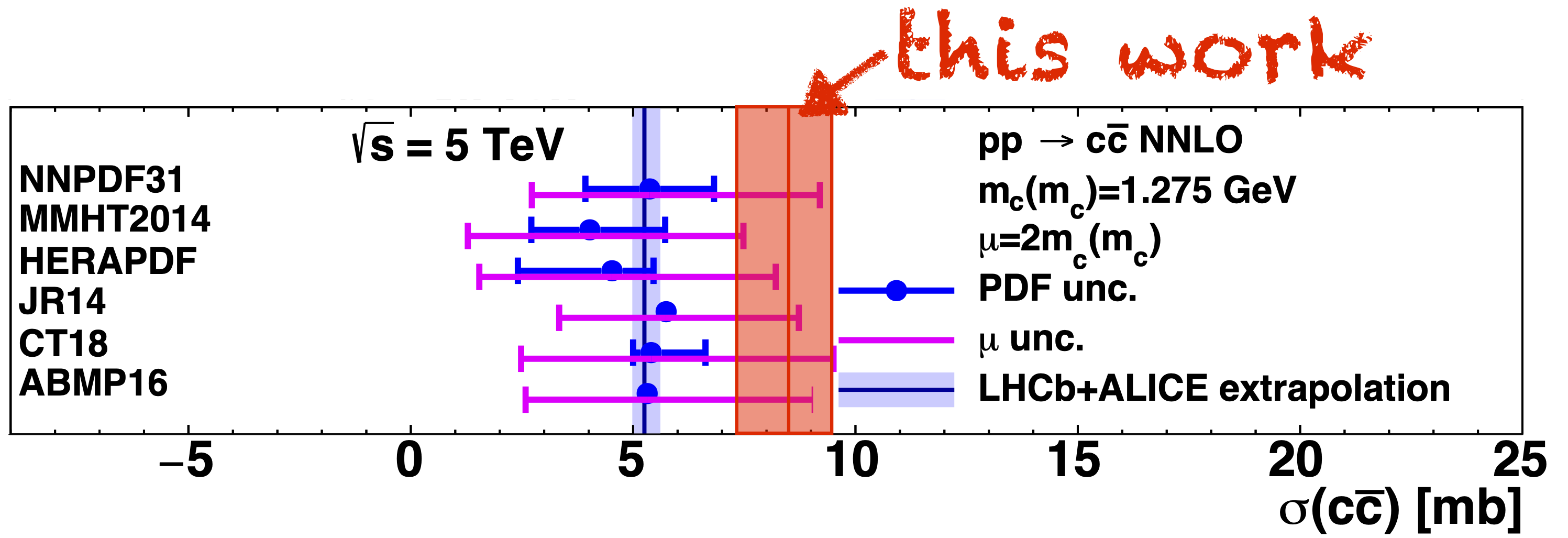}
 \end{center}
 \caption{Comparison of the total charm cross-section measurement at 5 TeV with NNLO theory. The figure is from \cite{nnloCharm1} and the result of this work accounting for the charm fragmentation non-universality is added (red band). 
The original result assuming universality is also shown (blue band). 
} \label{fig:nnloComp_5TeV}
\end{figure}

The total uncertainty for the non-universality case is increased relative to the one assuming universality since the measurement uncertainties of the fragmentation and production fractions in $pp$ data are larger than those in $e^+e^-$ data. Nevertheless, the total charm cross section obtained increases significantly. It is compared with NNLO predictions in Fig. \ref{fig:nnloComp_5TeV}. The measured total cross section still shows good agreement with the NNLO predictions, but prefers the upper edge of its uncertainty.

\section{Conclusion}

Recently LHC experiments have observed large differences of charm fragmentation between $pp$ and $e^+e^-/ep$ collisions, which are strongly related to a significant $p_T$ dependence of meson to baryon ratios. In this report, this non-universal charm fragmentation is applied and used to extrapolate $D^0$ measurements at 5 TeV from ALICE and LHCb in a novel phenomenological way which is independent of any particular non-universal non-perturbative fragmentation model. As a result, the total charm cross section is derived for the first time in a way that is consistent with measurements of charm fragmentation non-universality. It is demonstrated that the resulting cross section has increased compared to the one derived with the universality assumption, but remains consistent with NNLO theory, which is the most precisely known prediction for charm today. Since the extrapolation of the fiducial measurements was data driven, the measurement obtained is, within uncertainties, almost unbiased by theory. It can thus be directly used to further constrain QCD parameters such as PDFs or the charm quark mass. We acknowledge significant contributions to the initial studies by Max Uetrecht, and thank Sven-Olaf Moch and Oleksandr Zenaiev for feedback on the manuscript.

\bibliographystyle{unsrt}
\bibliography{eps23_cTotXsec5TeV}

\end{document}